\documentclass[conference]{IEEEtran}
\IEEEoverridecommandlockouts
\usepackage{cite}
\usepackage{amsmath,amssymb,amsfonts}
\usepackage{algorithmic}
\usepackage{graphicx}
\usepackage{textcomp}
\usepackage{color}
\usepackage{graphicx}
\usepackage{tikz}
\usepackage{color}
\usepackage{enumitem}
\usepackage{hyperref}

\usepackage{xcolor}
\def\BibTeX{{\rm B\kern-.05em{\sc i\kern-.025em b}\kern-.08em
    T\kern-.1667em\lower.7ex\hbox{E}\kern-.125emX}}
\begin{document}

\title{CelloAI Benchmarks: Toward Repeatable Evaluation of AI Assistants
\thanks{fmohammad@bnl.gov}
}


\author{
\IEEEauthorblockN{
    Mohammad Atif\IEEEauthorrefmark{1},
    Kriti Chopra\IEEEauthorrefmark{1},
    Fang-Ying Tsai\IEEEauthorrefmark{2},
    Ozgur O. Kilic\IEEEauthorrefmark{1}
    Tianle Wang\IEEEauthorrefmark{1},
    Zhihua Dong\IEEEauthorrefmark{1},\\
    Douglas Benjamin\IEEEauthorrefmark{1},
    Charles Leggett\IEEEauthorrefmark{3},
    Meifeng Lin\IEEEauthorrefmark{1},
    Paolo Calafiura\IEEEauthorrefmark{3},
    Salman Habib\IEEEauthorrefmark{4}
    }
    \IEEEauthorblockA{\IEEEauthorrefmark{1}Brookhaven National Laboratory, NY, USA \IEEEauthorrefmark{2}Stony Brook University, NY, USA}
    \IEEEauthorblockA{\IEEEauthorrefmark{3}Lawrence Berkeley National Laboratory, CA, USA \IEEEauthorrefmark{4}Argonne National Laboratory, IL, USA}
}

\maketitle

\begin{abstract}
Large Language Models (LLMs) are increasingly used for software development, yet existing benchmarks for LLM-based coding assistance do not reflect the constraints of High Energy Physics (HEP) and High Performance Computing (HPC) software.
Code correctness must respect science constraints and changes must integrate into large, performance-critical codebases with complex dependencies and build systems. 
This paper’s primary contribution is the development of practical, repeatable benchmarks that quantify LLM performance on HEP/HPC-relevant tasks.
We introduce three evaluation tracks: code documentation benchmarks measure an LLM's ability to generate Doxygen-style comments, code generation benchmarks evaluate end-to-end usability on representative GPU kernels, and graphical data analysis benchmarks evaluate vision-enabled LLMs.
These benchmarks provide a unified framework for measuring progress in scientific coding assistance across documentation quality, code generation robustness, and multimodal validation analysis. By emphasizing repeatability, automated scoring, and domain-relevant failure modes, the suite enables fair comparisons of models and settings while supporting future work on methods that improve reliability for HEP/HPC software development.
 \end{abstract}

\section{Introduction}

Recent advances in large language models (LLMs) have enabled new tools that can assist scientists with two  pain points in research software development cycle: (i) producing clear natural-language documentation for complex code, and (ii) generating or modifying code to accelerate development \cite{diggs2024leveraging,nam2024using,nejjar2025llms}. 
Yet, in high-performance computing (HPC) environments, practical adoption remains constrained by the difficulty of providing sufficient context for massive, sparsely documented codebases.  
Moreover, heterogeneous architectures amplify these challenges, where mistakes in translation, data movement, or dependency handling can silently invalidate results or degrade performance. 
CelloAI address these constraints as a locally hosted, retrieval-augmented coding assistant purpose-built for scientific HPC workflows \cite{atif2025celloai}.

CelloAI’s goal is to make LLM assistance reliable for large scientific repositories by assembling high-quality contexts before generation. The system is designed around three core ideas: (1) retrieval-augmented generation that pulls from both scientific text and source code; (2) syntax-aware code chunking to preserve semantic boundaries to avoid fragmented retrieval; and (3) callgraph-aware prompt enhancement to provide dependency context (caller/callee), so LLM responses respect the execution flow of real applications. Together, these mechanisms support documentation, code understanding, and code generation tasks while maintaining transparency and safety requirements that are essential in scientific computing. 


This paper expands CelloAI into a benchmark-driven framework spanning three HEP-relevant evaluation tracks with methodologies applicable to other HPC domains.
Benchmarking is essential because qualitative, anecdotal impressions do not provide a repeatable basis for comparing models or strategies across large codebases. 
Standardized tasks and automated scoring make performance claims testable, enable fair comparisons across systems, and reveal failure modes that are easy to miss in smaller studies. 
The main contribution of this paper is to provide practical strategies for developing benchmarks for code documentation, code generation/porting, and graphical data analysis, complementing existing CelloAI feaures.  
These benchmarks also provide a foundation for measuring progress in AI technologies and enabling fair comparisons between different models. 

Existing benchmarks are strong for general coding competence and unit test-driven software fixes \cite{jimenez2023swe,jain2024livecodebench,zhuo2024bigcodebench}, but they fall short of HEP/HPC needs where correctness requires scientific validation and performance \cite{austin2021program}.
Hence, we introduce benchmarks for \textbf{code documentation}, targeting structured documentation such as Doxygen-style comments, where success requires both complete coverage of parameters/returns and scientifically grounded descriptions. 
Second, we introduce benchmarks for \textbf{code generation and transformation} in HPC settings, emphasizing GPU porting where correctness, dependency awareness, and guardrailed behavior matter as much as raw generation ability. 
Third, we also extend CelloAI toward \textbf{graphical data analysis} benchmarks, leveraging histograms generated during HEP experiments and simulations. 
Our aim is to provide a unified and reproducible way to measure whether LLMs can generate code {accurately}, {completely}, {consistently}, and {with scientific grounding} under real HPC constraints.



\section{Code Documentation}

Code documentation is critical in HEP and HPC software because these code-bases are characteristically large and long-lived, developed by distributed teams, and must maintain understandability and reproducibility  despite decades of evolution with changing personnel and hardware. 
LLMs offer a practical way to automate parts of documentation by generating structured comments directly from function signatures and local code context, reducing the manual effort required to keep comments synchronized with rapidly changing implementations.
However, scientific documentation requires domain-specific terminology, making output quality dependent on whether models can accurately infer scientific intent from limited context. Additionally, output quality is also sensitive to instruction adherence, since formats such as Doxygen require consistent structure alongside meaningful descriptions. 
Our objective is therefore to determine which LLMs (and their associated parameters) most reliably produce high-quality Doxygen-style comments for scientific/HPC functions. 
A central challenge is that assessing descriptive text's correctness is inherently difficult.
Hence, we formalize benchmarks by combining structural measures of completeness with semantic measures that probe whether the generated text is coherent and aligned with expert intent.

\subsection{Doxygen-style Comment Generation with CelloAI}
CelloAI generates Doxygen-style comments by extracting code units and interface information, including signatures, argument names/types, and return types. 
It retrieves supporting context from the repository and linked scientific artifacts that describe the implemented algorithm or physics behavior. 
Using this context, the model produces syntactically valid Doxygen blocks with required tags  (e.g., \texttt{@brief}, \texttt{@param}, \texttt{@return}) and concise descriptions.

\subsection{CelloAI-Doc-Bench}
CelloAI-Doc-Bench is a lightweight benchmark suite to evaluate how well LLMs generate structured documentation for scientific and high-performance computing code.  
It computes two core metrics: a coverage score, measuring the fraction of documented parameters and return tags relative to the ground truth, and a semantic similarity score, measuring the coherency of generated text and and alignment with expert-written comments. Together these metrics capture both completeness and contextual grounding of the generated documentation.

\subsubsection{Coverage Score}
To evaluate structural correctness, we extend the standard F1 score [cite] to a tag-level coverage score in CelloAI-Doc-Bench.
The benchmark emphasizes coverage by ensuring that every parameter and return value in the function is reflected in \texttt{@param} and \texttt{@return} tags, and hence evaluates this with tag-level precision/recall and an $F_1$ summary.
We count the number of true positive (TP), false positive (FP), and false negative (FN) tags.
In this context, a TP occurs when the LLM correctly includes a tag for an existing parameter or return value; an FP occurs when it adds a tag that does not exist in the ground truth; and a FN occurs when it omits a tag that should be present.  
The final $F_1$ score is the harmonic mean of precision and recall scores defined as
\begin{align}
\begin{split}
F_1 &= \frac{2 \times \text{Precision} \times \text{Recall}}{\text{Precision} + \text{Recall}},\\ 
\text{Precision}&=\frac{TP}{TP+FP}, \text{Recall}=\frac{TP}{TP+FN}.        
\end{split}
\end{align}
Higher recall here means more complete documentation, while higher precision indicates fewer redundant tags, thus yielding an interpretable measure of documentation quality.

\subsubsection{Semantic Similarity}
\begin{figure}
\begin{tikzpicture}[scale=0.65, transform shape, font=\small]
\node[draw, align=left, inner sep=4pt] (caller) at (1,0) {%
\textbf{Caller}\\
\texttt{/**}\\
\texttt{ * @param param1 description ...}\\
\texttt{ * @param param2 description ...}\\
\texttt{ * @param param3 description ...}\\
\texttt{ */}\\
\texttt{void foo(dtype1 param1,}\\
\texttt{         dtype2 param2,}\\
\texttt{         dtype3 param3);}
};
\node[draw, align=left, inner sep=4pt] (callee) at (8.2,0) {%
\textbf{Callee}\\
\texttt{/**}\\
\texttt{ * @param param1 description ...}\\
\texttt{ * @param param22 description ...}\\
\texttt{ * @param param4 description ...}\\
\texttt{ */}\\
\texttt{void bar(dtype1 param1,}\\
\texttt{         dtype2 param22,;}\\
\texttt{         dtype4 param3);}
};
\draw[->] (caller.east) -- node[above]{call} (callee.west);
\node[draw, align=left, inner sep=4pt] (parse1) at (1.5,-3.0) {%
Parse Doxygen\\
-- extract params and descriptions
};
\node[draw, align=left, inner sep=4pt] (parse2) at (7.5,-3.0) {%
Parse Doxygen\\
-- extract params and descriptions
};
\draw[->] (caller.south west) -- (parse1.north west);
\draw[->] (callee.south east) -- (parse2.north east);
\node[draw, align=left, inner sep=4pt] (match) at (4.5,-5) {%
Name match filter\\
-- select only if types and names equal\\
\texttt{dtype1 param1} = \texttt{dtype1 param1}
};
\draw[->] (parse1.south west) -- (match.west);
\draw[->] (parse2.south east) -- (match.east);
\node[draw, align=left, inner sep=4pt] (sim) at (4.5,-6.7) {%
Embed + compute cosine similarity between descriptions of \texttt{param1}
};
\draw[->] (match.south) -- (sim.north);
\end{tikzpicture}
\caption{Differential Similarity Schematic: Parameters whose names are identical in both the caller and the callee indicate an intentional reuse, thus should be semantically similar.}
\label{fig:diffsim}
\end{figure}
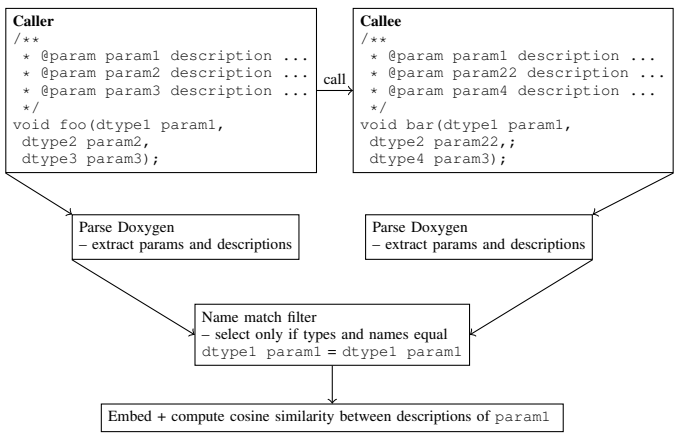

While tag-based coverage score evaluates structural accuracy, it does not capture whether generated documentation is meaningful.
To address this, we introduce two semantic similarity scores using cosine similarity between vector representations of descriptive text:
\begin{itemize}[leftmargin=*]
\item Differential Similarity:Coherent LLM documentation should show consistent parameter descriptions across related functions. 
This benchmark evaluates semantic consistency of parameter descriptions across caller–callee function pairs by comparing their Doxygen comments.    
Since parameter names are locally scoped and may be renamed, we restrict analysis to parameters with identical names in both caller and callee, indicating intentional semantic reuse.
 This schematic is represented in Figure \ref{fig:diffsim}. 
 This constraint yields a conservative but high-precision evaluation set. 
 We apply this benchmark FastCaloSim's CUDA port, a parametrized simulation of ATLAS Liquid Argon Calorimeter \cite{hasib2017new}, identifying 18 caller–callee parameter pairs sharing the same argument name. 
 For each such pair, we compute the cosine similarity between dense embedding representations of the corresponding parameter descriptions. 
 Finally, the Differential Similarity Score is the mean semantic similarity across all matched parameter pairs, providing a quantitative measure of documentation coherence. 
\item Expert Similarity: To evaluate whether automated documentation approaches expert quality, we compare LLM-generated comments against expert-written documentation.
We select Calorimeter/CaloClusterCorrection/src directory from ATLAS' Athena package \cite{athena_calo_subset} and extract all functions and their corresponding expert-written comments. 
The LLM is then prompted to generate Doxygen-style comment.
This is followed by generating vector representations of text using an embedding LLM and then measuring the cosine of angle between the vectors of the expert comment and LLM-generated comment.
Finally, the Expert Similarity is calculated as the average of cosines between all extracted comments.
Thus, this metric captures the semantic proximity between LLM and expert documentation.
\end{itemize}

\begin{table*}[]
    \centering
    \begin{tabular}{c|c|c|c|c|c}
        & \multicolumn{3}{c|}{Tag-based Coverage} &  \multicolumn{2}{c}{Semantic Similarity} \\
        Model/Parameter & Precision & Recall & F1 & Differential & Expert \\
        \hline \\
Meta/Llama-3-7B ($\theta=0.1$)             & 0.175 (0.063) & 0.032 (0.021) & 0.051 (0.026) & 0.000 (0.000) & 0.471 (0.006) \\        
Meta/Llama-3.2-3B ($\theta=0.1$)	          & 0.881 (0.033) & \textbf{1.000} (0.000) & 0.927 (0.023) & \textbf{0.677} (0.019) & 0.579 (0.002) \\
Microsoft/phi4 ($\theta=0.1$)	      & 0.902 (0.016) & 0.987 (0.004) & 0.934 (0.013) & 0.647 (0.025) & 0.575 (0.001) \\
\hline
OpenAI/Gpt-oss-120b ($\theta=0.1$)	          & 0.918 (0.019) & 0.998 (0.005) & 0.952 (0.011) & 0.657 (0.023) &	0.602 (0.002) \\
OpenAI/Gpt-oss-120b ($\theta=0.5$)	          &	0.922 (0.014) & \textbf{1.000} (0.000) & 0.956 (0.008) & 0.643 (0.017) & 0.600 (0.003) \\
OpenAI/Gpt-oss-120b ($\theta=1.0$)	          &	0.931 (0.012) & 0.999 (0.001) & 0.960 (0.007) & 0.633 (0.015) & 0.581 (0.006) \\
\hline
OpenAI/Gpt-oss-120b ($\theta=0.1$, CelloAI ) & \textbf{0.942} (0.006) & \textbf{1.000} (0.000) & \textbf{0.967} (0.004) & 0.656 (0.029) & \textbf{0.619} (0.003) \\
OpenAI/Gpt-oss-120b ($\theta=0.5$, CelloAI ) &	0.935 (0.011) & \textbf{1.000} (0.000) & 0.963 (0.006) & 0.648 (0.024) & 0.604 (0.003) \\
OpenAI/Gpt-oss-120b ($\theta=1.0$, CelloAI ) &	0.933 (0.006) & 0.997 (0.007) & 0.960 (0.005) & 0.651 (0.027) & 0.601 (0.004) \\
\hline
Qwen/Qwen3-8b ($\theta=0.1$)               & 0.864 (0.033) & \textbf{1.000} (0.000) & 0.908 (0.029) & 0.588 (0.045) & 0.570 (0.009) \\
Qwen/Qwen3-8b ($\theta=0.5$)		          & 0.902 (0.012) & \textbf{1.000} (0.000) & 0.937 (0.010) & 0.590 (0.030) & 0.568 (0.007) \\
    \end{tabular}
    \caption{CelloAI-Doc-Bench scores across models and LLM temperatures $\theta$: tag-coverage precision/recall/F1 and semantic similarity scores averaged over 10 runs with standard deviations in parentheses}
    \label{tab:code-doc-bench}
\end{table*}

Table \ref{tab:code-doc-bench} depicts CelloAI-Doc-Bench scores across LLMs (and their generation temperatures $\theta$). 
CelloAI's features (callgraph, patterns matching, and re-ranking) themselves are considered parameters of the LLM for the case when CelloAI context workflow is enabled.
Each row reports one model and all the scores (higher is better) are averaged over 10 runs.
The three tag-based coverage columns report the precision/recall/F1 scores in the Doxygen tags.
The two semantic similarity columns measure the differential and expert similarity of the comments.
Most new LLMs achieve very high recall ($\approx 1.0$) and high F1 ($\approx 0.91 - 0.97$), meaning they usually include the full set of required \texttt{@param}/\texttt{@return} tags. 
The older models like Llama-3-7B at and Llama-3.2-3B have very low recall and F1, indicating frequent missing tags. 
For semantic quality, the best Expert Similarity scores are obtained by the larger model configurations (up to ($\sim 0.62$)), while smaller models tend to be lower ($(\sim 0.57) - 0.58$) even with strong tag coverage. 
Across $\theta$ for gpt-oss-120b, tag coverage changes little, while semantic similarity tends to decrease slightly at higher $\theta$, suggesting that more stochastic decoding can reduce description consistency and alignment with experts.
We also find that semantic similarity remains limited (likely requiring domain fine-tuning), and enabling CelloAI yields only marginal changes in these metrics.

\section{HPC Code Generation}

Recently, AI coding tools such as GitHub Copilot, Google’s Gemini Code Assist, and Anthropic’s Claude Code have emerged, primarily designed for interactive development workflows.
In parallel, widely used evaluation suites for code generation and repair (e.g., SWE-bench, LiveCodeBench, and BigCodeBench) emphasize unit-test passing and problem-solving in comparatively self-contained settings \cite{jimenez2023swe,jain2024livecodebench,zhuo2024bigcodebench,qiu2024efficient}.
For HPC applications, code generation tasks often involve translating algorithms across architectures, refactoring kernels for performance portability, or implementing small but critical components that interact with experiment-specific data structures \cite{ouyang2025kernelbench,atif2026evaluating}. 
These tasks are error-prone when attempted from local context alone, because correct implementations typically depend on surrounding conventions (memory layouts, data ownership, and interface contracts) and on assumptions documented outside the code (notes, slides, or prior analyses). 
In HEP, correctness includes science constraints (physics intent, numerical stability) and system constraints (CPU-GPU transfers, memory layout, performance-portability) not captured by typical unit tests. 
Consequently, HEP/HPC code generation benchmarks should prioritize end-to-end outcomes – compilation, execution, and domain validation rather than isolated unit-tests.



\subsection{Code Porting with CelloAI}

CelloAI code generation is designed to produce and modify code with stronger reliability by grounding generation in retrieved RAG context and dependency information. 
Code porting must focus on translating existing scientific kernels to new environments (e.g., CUDA to OpenMP or SYCL) while preserving correctness and expected performance behavior \cite{lin2023portable}. 
Hence, CelloAI  uses retrieval over the target repository to surface relevant patterns and local conventions so the generated port matches project style and integrates cleanly. 
To avoid local fixes that break downstream code, CelloAI incorporates dependency context (caller - callee relationships) so changes respect function interfaces and usage. 

\subsection{CelloAI-Code-Bench}

CelloAI-Code-Bench is a suite of benchmarks being developed where correctness and integration into an existing codebase matter more than standalone snippet generation. 
Firstly, we leverage ATLAS FastCaloSim's \cite{atif2024porting} CUDA version to generate more portable OpenMP GPU code.
FastCaloSim’s workflow consists of three GPU kernels: 
\begin{itemize}[leftmargin=*]
    \item \textbf{Reset} kernel clears a large per-event device array representing the cells of a detector
    \item The \textbf{count} kernel identifies hit cells and packs results for efficient transfer back to the host.
    \item The compute-heavy \textbf{simulation} kernel  applies the selected shower parameterizations to deposit energy into detector elements. It relies on atomics for accumulation. 
\end{itemize}
In practice, the simulation kernel is the hardest as it mixes heavy floating-point computations with memory updates and  host-to-device data transfers. 
In this kernel, any missing mappings will break correctness. 
The count kernel is moderate as the LLM has to correctly identify an atomic operation, while the reset kernel is easiest at it only requires setting device arrays to zero. 

The purpose of the three-kernel benchmark is to provide a simple, repeatable way to measure whether an LLM-assisted porting workflow can produce working GPU implementations for representative FastCaloSim components. 
Each benchmark trial presents the model with the same kernel-level task under a fixed context condition (e.g., baseline retrieval versus CelloAI’s enhanced context). 
The model’s output is then evaluated by an automated pipeline that attempts to integrate the generated code, compile it, and run a predefined validation step; a trial is counted as successful only if it passes these checks.

\begin{table*}[]
    \centering
    \begin{tabular}{c|c|c|c}
    Model/Parameter & Kernel 1 [reset] & Kernel 2 [count] & Kernel 3 [simulate] \\
    \hline
    OpenAI/Gpt-oss-120b ($\theta=0.1$, RAG) &  2/10 & 3/10 & 0/10 \\
    OpenAI/Gpt-oss-120b ($\theta=0.5$, RAG) &  4/10 & 2/10 & 0/10 \\
    OpenAI/Gpt-oss-120b ($\theta=1.0$, RAG) &  5/10 & 5/10 & 0/10 \\
    \hline
    OpenAI/Gpt-oss-120b ($\theta=0.1$, CelloAI) &  9/10  & 3/10 & 1/10 \\
    OpenAI/Gpt-oss-120b ($\theta=0.5$, CelloAI) &  9/10  & \textbf{8/10} & 1/10 \\
    OpenAI/Gpt-oss-120b ($\theta=1.0$, CelloAI) &  \textbf{10/10} & 6/10 & \textbf{2/10} \\ 
    \hline
    MistralAI/Mistral-large-2407 ($\theta=0.1$, RAG) &  3/10 & 0/10 & 0/10 \\
    MistralAI/Mistral-large-2407 ($\theta=0.5$, RAG) &  2/10 & 0/10 & 0/10 \\
    MistralAI/Mistral-large-2407 ($\theta=1.0$, RAG) &  1/10 & 0/10 & 0/10 \\
    \hline
    Qwen/Qwen3-8b ($\theta=1.0$, RAG) & 0/10 & 0/10 & 0/10 \\
    Microsoft/phi4 ($\theta=1.0$, RAG) & 0/10 &  0/10 & 0/10 \\
    OpenAI/Gpt-oss-20b ($\theta=1.0$, RAG)    & 6/10 &  1/10 & 0/10 \\
    \end{tabular}
    \caption{Automated code-generation benchmark results for three FastCaloSim GPU kernels. Each entry reports the number of successful runs out of 10 (compile + validation pass) for a given model and temperature $\theta$ under a specified context condition (RAG baseline or CelloAI).}
    \label{tab:codebench}
\end{table*}

In Table \ref{tab:codebench}, we compare LLMs by counting how many kernels can be ported to a correct, runnable implementation under a fixed number of attempts, and briefly compare retrieval-only baselines against CelloAI’s enhanced, context-aware generation.
Each row reports one model and decoding setting (temperature $\theta$ and CelloAI).
We observe that across all configurations, Kernel 3 (simulate) is the most difficult: most models achieve 0 successes, and even the best results with Gpt-oss-120b reach only 2/10 indicating that end-to-end correctness for the main simulation kernel is rarely achieved. 
For Gpt-oss-120b under the baseline retrieval setting (RAG), performance on Kernel 1 and Kernel 2 improves modestly as $\theta$ increases.
The Gpt-oss-120b runs with the CelloAI configuration show higher success on Kernel 1 (9–10/10) and improved performance on Kernel 2, while Kernel 3 remains low (1–2/10), reinforcing that the simulate kernel dominates overall failure rates. 
Other models (Mistral-large-2407, Qwen3-8b, Microsoft/phi4) do not succeed on Kernel 2 or 3 in these settings, and only show limited success on Kernel 1.
Thus, this benchmark separates simple kernels from the scientifically and structurally constrained simulation kernel.

\section{Graphical Data Analysis}
Graphical data analysis is a core part of scientific software development. 
Traditional computer-vision tools (e.g., OpenCV) and web plot digitizers can detect basic features or extract numerical traces from images, but they do not capture the scientific intent of a plot or support higher-level interpretation. 
In many scientific workflows, analysts review large batches of plots and must relate observed changes to specific algorithms, configurations, or code paths in a complex software stack. 
Making these connections requires visual pattern recognition \cite{methani2020plotqa,masry2022chartqa}, reasoning about what constitutes a meaningful deviation, and interpretation in the context of the underlying codebase. 
Vision-enabled LLMs can analyze rendered graphs directly to flag anomalies, describe the type of change, and prioritize the most informative plots for follow-up.

\subsection{Histogram Analysis with CelloAI}
Reconstruction, simulation, and calibration workflows in HEP routinely produce large collections of histograms spanning many physics objects, detector regions, and selections.
These histograms are compared across software versions to detect unexpected changes. 
When differences appear, the key problem is not only detection, but interpretation, i.e., deciding whether a change indicates a physics regression, a benign numerical shift, or an intended algorithm update, and then tracing the most likely origin within a large and evolving codebase. 
In CelloAI, we are developing this connection between histogram analysis and software context by combining vision-based interpretation of histogram deviations with retrieval of  code knowledge (e.g., callgraph and documentation metadata) to suggest which modules or routines could plausibly explain the observed pattern. 
This motivates graphical data analysis as a benchmark that evaluates multimodal capabilities of vision LLMs.

\subsection{CelloAI-Multimodal-Bench}

\begin{figure}
    \centering
    \includegraphics[width=0.8\linewidth]{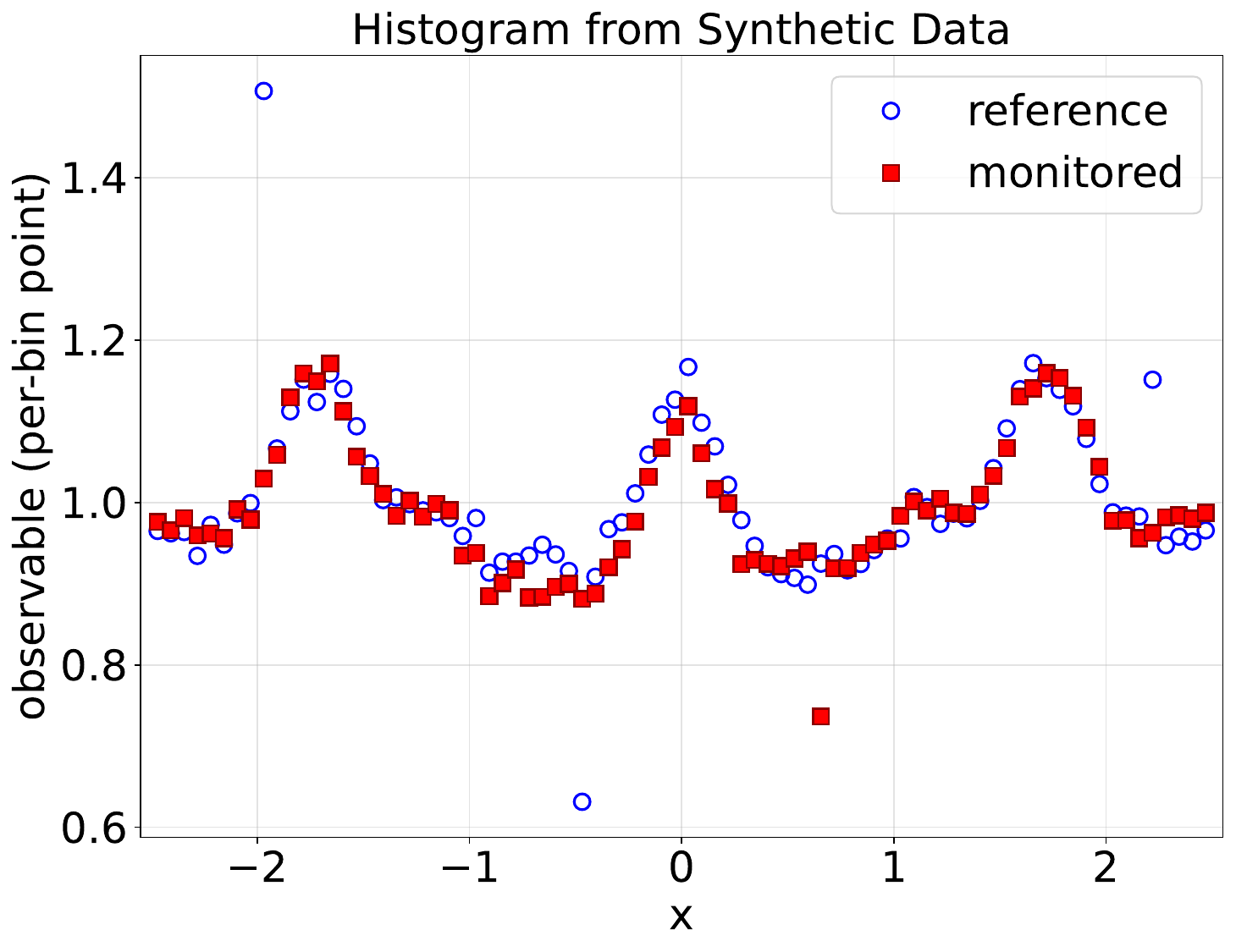}
    \caption{Synthetic histogram typically used in HEP workflows.}
    \label{fig:synthist}
\end{figure}

CelloAI-Multimodal-Bench is a multimodal suite of benchmarks under development.
In this benchmark, we use a synthetic histogram (see Figure \ref{fig:synthist}) to provide ground truth, and evaluate whether a vision model can (i) detect the discrepancy between `reference' and `monitored' curves, and (ii) list all the outliers. 
Here, each LLM outputs JSON with outlier points and discrepancy-region intervals.
For outliers, the benchmark counts a prediction correct if it matches a ground-truth point within a specified tolerance, and then computes precision/recall/F1 scores.
For discrepancies, each predicted discrepancy interval is converted into a set of locations by sampling the x-axis every 0.05 (1.25x the bin width), which is then compared with ground-truth's samples to again compute precision/recall/F1.

\begin{table*}[]
    \centering
    \begin{tabular}{c|c|c|c|c|c|c}
     & \multicolumn{3}{c|}{Outlier} & \multicolumn{3}{c}{Discrepancy} \\
    Model/Parameter & Precision & Recall & F1 & Precision & Recall & F1 \\
    \hline
Qwen/Qwen3-VL-8B-Instruct ($\theta=0.1$) & 0.5 (0)	& \textbf{0.5} (0) & 0.5 (0) & \textbf{0.548} (0)	    & \textbf{0.657} (0)	    & \textbf{0.597} (0) \\
Qwen/Qwen3-VL-8B-Instruct ($\theta=0.5$) & 0.5 (0)	& \textbf{0.5} (0) & 0.5 (0)	& \textbf{0.548} (0)	    & \textbf{0.657} (0)	    & \textbf{0.597} (0) \\
Qwen/Qwen3-VL-8B-Instruct ($\theta=1.0$) & 0.5 (0)	& \textbf{0.5} (0) & 0.5 (0)	& 0.423 (0.226)	& 0.386 (0.267)	& 0.39 (0.237) \\
\hline					
Google/Gemma-3n-E4B-it ($\theta=0.1$) & 0 (0) & 0 (0) & 0 (0) & \textbf{0.548} (0)	  & \textbf{0.657} (0)	   & \textbf{0.597} (0) \\
Google/Gemma-3n-E4B-it ($\theta=0.5$) & 0 (0) & 0 (0) & 0 (0) & 0.341 (0.182) & 0.431 (0.208) & 0.38 (0.194) \\
Google/Gemma-3n-E4B-it ($\theta=1.0$) & 0 (0) & 0 (0) & 0 (0) & 0.244 (0.133) & 0.291 (0.155) & 0.26 (0.135) \\
\hline
OpenGVLab/InternVL 3.5 8B ($\theta=0.1$) & \textbf{0.667} (0)	   & \textbf{0.5} (0)    	 & \textbf{0.571} (0)	& 0.218 (0.070)	& 0.137 (0.0439) & 0.169 (0.0547) \\
OpenGVLab/InternVL 3.5 8B ($\theta=0.5$) & \textbf{0.667} (0)	   & \textbf{0.5} (0)	     & \textbf{0.571} (0)	& 0.289 (0.204)	& 0.327 (0.225)	 & 0.304 (0.228) \\
OpenGVLab/InternVL 3.5 8B ($\theta=1.0$) & 0.448 (0.312) & 0.344 (0.229) &0.388 (0.264)	& 0.307 (0.218)	& 0.275 (0.207)	 & 0.288 (0.212) \\
    \end{tabular}
    \caption{Precision, recall, and F1 for outlier detection and discrepancy-region identification across models and temperature $\theta$. Values report the mean over 10 independent runs, with the standard deviation shown in parentheses.}
    \label{tab:multimodal}
\end{table*}

Table \ref{tab:multimodal} shows a comparison of three vision LLMs at various temperatures $\theta$. We observe that outlier detection is strongest for InternVL at low-to-moderate temperatures (0.1 and 0.5), achieving an F1 of 0.571, while Qwen3-VL is consistently moderate with F1 fixed at 0.5 across all tested settings. 
In contrast, Gemma-3n fails to recover any ground-truth outliers.
For discrepancy-region identification, Qwen3-VL at 0.1 and 0.5 attains the best overall performance, with Gemma-3n matching this level only at the lowest $\theta$ but degrading substantially as temperature increases. 
Qwen3-VL similarly deteriorates at $\theta=1.0$, whereas InternVL remains comparatively weak on discrepancy detection across $\theta$, improving first from $\theta$ 0.1 to 0.5 and then plateauing.
Overall, the scores remain relatively moderate across tasks and settings, indicating that more capable or domain-adapted (fine-tuned) multimodal models are likely needed to reliably extract outliers and discrepancy regions from scientific plots.
    
\section{Conclusion and Outlook}

This work presents a benchmark-driven framework for evaluating LLM capabilities on HEP/HPC-relevant tasks. 
Rather than relying on anecdotal impressions, we define repeatable tasks with automated scoring for three  areas: (i) Doxygen-style code documentation, (ii) kernel-level code generation and porting for FastCaloSim, and (iii) vision-enabled graphical data analysis on synthetic data. The benchmarks reveal successes and expose failure modes. 
For documentation, many models achieve strong structural coverage while semantic quality remains limited, indicating a need for better domain grounding. 
For code generation, success rates depend strongly on kernel complexity: simpler kernels are frequently solvable while the main simulation kernel remains difficult, emphasizing that compilation and validation should be primary evaluation criteria in scientific settings.
For graphical data analysis, we outline a multimodal benchmark coupling deviations with automated outlier detection, addressing real-world requirements for understanding differences across large histogram collections.

A key outcome of this paper is a methodology for real-world evaluation of emerging AI technologies in scientific software development. 
The benchmark definitions and scoring procedures are designed to be reproducible and to support fair comparison across models, decoding settings, and future methods. 
In future, we will expand coverage to additional open and closed LLMs, include more fine-tuned and domain-adapted models, and broaden the set of tasks and repositories to better represent the diversity of HEP workloads. 
Overall, the benchmark suite provides a foundation for measuring progress toward reliable, science-aware code assistants and  guiding the development of systems that operate under the constraints of large, performance-critical scientific codebases.

\section*{Acknowledgment}

This work is supported by US Department of Energy, Office of Science, Office of High Energy Physics under the High Energy Physics Center for Computational Excellence (HEP-CCE) under B\&R KA2401045, a collaboration between Argonne National Laboratory, Brookhaven National Laboratory, Fermilab, Oak Ridge National Laboratory, and Lawrence Berkeley National Laboratory.   

\bibliographystyle{IEEEtran}  
\bibliography{refs}

@inproceedings{nam2024using,
  title={Using an {LLM} to help with code understanding},
  author={Nam, Daye and Macvean, Andrew and Hellendoorn, Vincent and Vasilescu, Bogdan and Myers, Brad},
  booktitle={Proceedings of the IEEE/ACM 46th International Conference on Software Engineering},
  pages={1--13},
  year={2024}
}

@article{diggs2024leveraging,
  title={Leveraging {LLM}s for legacy code modernization: Challenges and opportunities for {LLM}-generated documentation},
  author={Diggs, Colin and Doyle, Michael and Madan, Amit and Scott, Siggy and Escamilla, Emily and Zimmer, Jacob and Nekoo, Naveed and Ursino, Paul and Bartholf, Michael and Robin, Zachary and others},
  journal={arXiv preprint arXiv:2411.14971},
  year={2024}
}

@article{nejjar2025llms,
  title={{LLM} for science: Usage for code generation and data analysis},
  author={Nejjar, Mohamed and Zacharias, Luca and Stiehle, Fabian and Weber, Ingo},
  journal={Journal of Software: Evolution and Process},
  volume={37},
  number={1},
  pages={e2723},
  year={2025},
  publisher={Wiley Online Library}
}

@article{atif2025celloai,
  title={Cello{AI}: Leveraging Large Language Models for HPC Software Development in High Energy Physics},
  author={Atif, Mohammad and Chopra, Kriti and Kilic, Ozgur and Wang, Tianle and Dong, Zhihua and Leggett, Charles and Lin, Meifeng and Calafiura, Paolo and Habib, Salman},
  journal={arXiv preprint arXiv:2508.16713},
  year={2025}
}

@softwareversion {athena_calo_subset,
  title = {Athena},
  author = {ATLAS},
  url = {https://gitlab.cern.ch/atlas/athena/-/tree/main/Calorimeter/CaloClusterCorrection/src?ref_type=heads},
}

@article{jain2024livecodebench,
  title={Livecodebench: Holistic and contamination free evaluation of large language models for code},
  author={Jain, Naman and Han, King and Gu, Alex and Li, Wen-Ding and Yan, Fanjia and Zhang, Tianjun and Wang, Sida and Solar-Lezama, Armando and Sen, Koushik and Stoica, Ion},
  journal={arXiv preprint arXiv:2403.07974},
  year={2024}
}

@article{jimenez2023swe,
  title={Swe-bench: Can language models resolve real-world github issues?},
  author={Jimenez, Carlos E and Yang, John and Wettig, Alexander and Yao, Shunyu and Pei, Kexin and Press, Ofir and Narasimhan, Karthik},
  journal={arXiv preprint arXiv:2310.06770},
  year={2023}
}

@article{zhuo2024bigcodebench,
  title={Bigcodebench: Benchmarking code generation with diverse function calls and complex instructions},
  author={Zhuo, Terry Yue and Vu, Minh Chien and Chim, Jenny and Hu, Han and Yu, Wenhao and Widyasari, Ratnadira and Yusuf, Imam Nur Bani and Zhan, Haolan and He, Junda and Paul, Indraneil and others},
  journal={arXiv preprint arXiv:2406.15877},
  year={2024}
}

@article{lin2023portable,
  title={Portable Programming Model Exploration for {L}Ar{TPC} Simulation in a Heterogeneous Computing Environment: {O}pen{MP} vs. {S}YCL},
  author={Lin, Meifeng and Dong, Zhihua and Wang, Tianle and Atif, Mohammad and Bhattacharya, Meghna and Knoepfel, Kyle and Leggett, Charles and Viren, Brett and Yu, Haiwang},
  journal={arXiv preprint arXiv:2304.01841},
  year={2023}
}

@article{atif2026evaluating,
  title={Evaluating Application Characteristics for {GPU} Portability Layer Selection},
  author={Atif, Mohammad and Bhattacharya, Meghna and Dewing, Mark and Dong, Zhihua and Esseiva, Julien and Gutsche, Oliver and Kortelainen, Matti and Kwok, Ka Hei Martin and Leggett, Charles and Lin, Meifeng and Strelchenko, Aleksei and Tsulaia, Vakhang and Viren, Brett and Wang, Tianle and Yu, Haiwang},
  journal={arXiv preprint arXiv:2601.17526},
  year={2026}
}

@inproceedings{atif2024porting,
  title={Porting {ATLAS} Fast Calorimeter Simulation to {GPU}s with Performance Portable Programming Models},
  author={Atif, Mohammad and Dong, Zhihua and Leggett, Charles and Lin, Meifeng and Tsulaia, Vakhtang},
  booktitle={EPJ Web of Conferences},
  volume={295},
  pages={11018},
  year={2024},
  organization={EDP Sciences}
}

@techreport{hasib2017new,
  title={The new ATLAS Fast Calorimeter Simulation},
  author={Hasib, Ahmed},
  year={2017},
  institution={ATL-COM-SOFT-2017-036}
}

@article{qiu2024efficient,
  title={How efficient is llm-generated code? a rigorous \& high-standard benchmark},
  author={Qiu, Ruizhong and Zeng, Weiliang Will and Ezick, James and Lott, Christopher and Tong, Hanghang},
  journal={arXiv preprint arXiv:2406.06647},
  year={2024}
}

@article{austin2021program,
  title={Program synthesis with large language models},
  author={Austin, Jacob and Odena, Augustus and Nye, Maxwell and Bosma, Maarten and Michalewski, Henryk and Dohan, David and Jiang, Ellen and Cai, Carrie and Terry, Michael and Le, Quoc and others},
  journal={arXiv preprint arXiv:2108.07732},
  year={2021}
}

@article{ouyang2025kernelbench,
  title={Kernelbench: Can llms write efficient gpu kernels?},
  author={Ouyang, Anne and Guo, Simon and Arora, Simran and Zhang, Alex L and Hu, William and R{\'e}, Christopher and Mirhoseini, Azalia},
  journal={arXiv preprint arXiv:2502.10517},
  year={2025}
}

@inproceedings{masry2022chartqa,
  title={Chartqa: A benchmark for question answering about charts with visual and logical reasoning},
  author={Masry, Ahmed and Do, Xuan Long and Tan, Jia Qing and Joty, Shafiq and Hoque, Enamul},
  booktitle={Findings of the association for computational linguistics: ACL 2022},
  pages={2263--2279},
  year={2022}
}

@inproceedings{methani2020plotqa,
  title={Plotqa: Reasoning over scientific plots},
  author={Methani, Nitesh and Ganguly, Pritha and Khapra, Mitesh M and Kumar, Pratyush},
  booktitle={Proceedings of the ieee/cvf winter conference on applications of computer vision},
  pages={1527--1536},
  year={2020}
}

\end{document}